\documentclass[english,prl,twocolumn,superscriptaddress]{revtex4}
\usepackage{times}
\usepackage[T1]{fontenc}
\usepackage[latin1]{inputenc}
\usepackage{amsmath}
\usepackage{color}
\usepackage{graphicx}
\usepackage{amssymb}
\makeatletter
\usepackage{babel}
\makeatother
\begin{document}

\title{Observation of Giant Positive Magnetoresistance in a Cooper Pair Insulator}

\date{\today{}}

\author{H. Q. Nguyen}
\altaffiliation{These authors contributed equally to this work.}
\author{S. M. Hollen}
\altaffiliation{These authors contributed equally to this work.}
\author{M. D. Stewart, Jr.}
\affiliation{Department of Physics, Brown University, Providence, RI 02912}
\author{J. Shainline}
\affiliation{Department of Physics, Brown University, Providence, RI 02912}
\author{Aijun Yin}
\affiliation{Division of Engineering, Brown University, Providence, RI 02912}
\author{J. M. Xu}
\affiliation{Division of Engineering, Brown University, Providence, RI 02912}
\author{J. M. Valles, Jr.}
\affiliation{Department of Physics, Brown University, Providence, RI 02912}

\begin{abstract}
Ultrathin amorphous Bi films, patterned with a nano-honeycomb array of holes, can exhibit an insulating phase with transport dominated by the incoherent motion of Cooper Pairs (CP) of electrons between localized states.  Here we show that the magnetoresistance (MR) of this Cooper Pair Insulator (CPI) phase is positive and grows exponentially with decreasing temperature, $T$, for $T$ well below the pair formation temperature. It peaks at a field estimated to be sufficient to break the pairs and then decreases monotonically into a regime in which the film resistance assumes the $T$ dependence appropriate for weakly localized single electron transport.  We discuss how these results support proposals that the large MR peaks in other unpatterned, ultrathin film systems disclose a CPI phase and provide new insight into the CP localization.    
 
\end{abstract}
\maketitle
 
Below its transition temperature, $T_{c0}$, a conventional superconductor, like Pb, can be driven into its non-superconducting, normal state by applying a magnetic field, $H$.  The temperature, $T$, dependent sheet resistance, $R_\square(T)$, of this state joins smoothly to the normal state resistance just above $T_{c0}$, $R_N$, and assumes the dependence expected for a simple metal of unpaired electrons\cite{Tinkham:Intro}.   The behavior of this low $T$ normal state changes substantially when these superconductors are made as thin films with $R_N\sim R_Q=h/4e^2$.  For example, applying a $H$ to superconducting (SC) films of either Indium oxide (InO$_x$)\cite{Hebard:PRL1990,Gantmakher:JETP1998,Sambandamurthy:PRL2004} or TiN\cite{Baturina:PRL2007} drives them directly into an insulating phase.  Their $R_\square(T)$ rise exponentially with decreasing $T$ to exceed $R_N$ by orders of magnitude.  A similarly dramatic Superconductor to Insulator Transition (SIT) can be achieved in granular Pb films by increasing $R_N$\cite{Barber:PRB1994}.  These behaviors have been taken to imply that the films enter a Cooper Pair Insulator (CPI) phase consisting of exponentially localized, but intact Cooper Pairs of electrons\cite{Barber:PRB1994,Gantmakher:JETP1998,Stewart:Science2007,Stewart:PRB2008}.   

For granular Pb films, the formation of a CPI phase has strong intuitive appeal and experimental support.  STM experiments show that they consist of islands of grains that can naturally localize CPs\cite{Ekinci:PRL1999}.  Indeed, tunneling experiments on insulating films confirmed the existence of these localized pairs by showing the energy gap in the density of states that accompanies CP formation\cite{Barber:PRB1994}.  Also, these insulators exhibit giant negative Magneto-Resistance (MR) that can be attributed to the enhancement of inter-island quasi-particle tunneling\cite{Barber:PRB2006}.   By contrast, InO$_x$ and TiN films lack any obvious structure that could localize CPs.  Moreover,  these films exhibit a giant positive MR\cite{Gantmakher:JETP1998,Sambandamurthy:PRL2004,Steiner:PhysicaC2005,Tan:PRB2008,Baturina:PRL2007}, which can peak orders of magnitude above $R_N$ at sufficiently low $T$.  The mechanism behind this spectacular giant positive MR\cite{Gantmakher:JETP1998,Sambandamurthy:PRL2004,Baturina:PRL2007,Steiner:PhysicaC2005,Tan:PRB2008} and whether it is a property of a CPI phase remains unresolved despite significant attention\cite{Fisher:PRL1990,Galitski:PRL2005,Dubi:PRB2006,Fistul:PRL2008}. 
 
 \begin{figure}
\begin{center}
\includegraphics[width=\columnwidth,keepaspectratio]{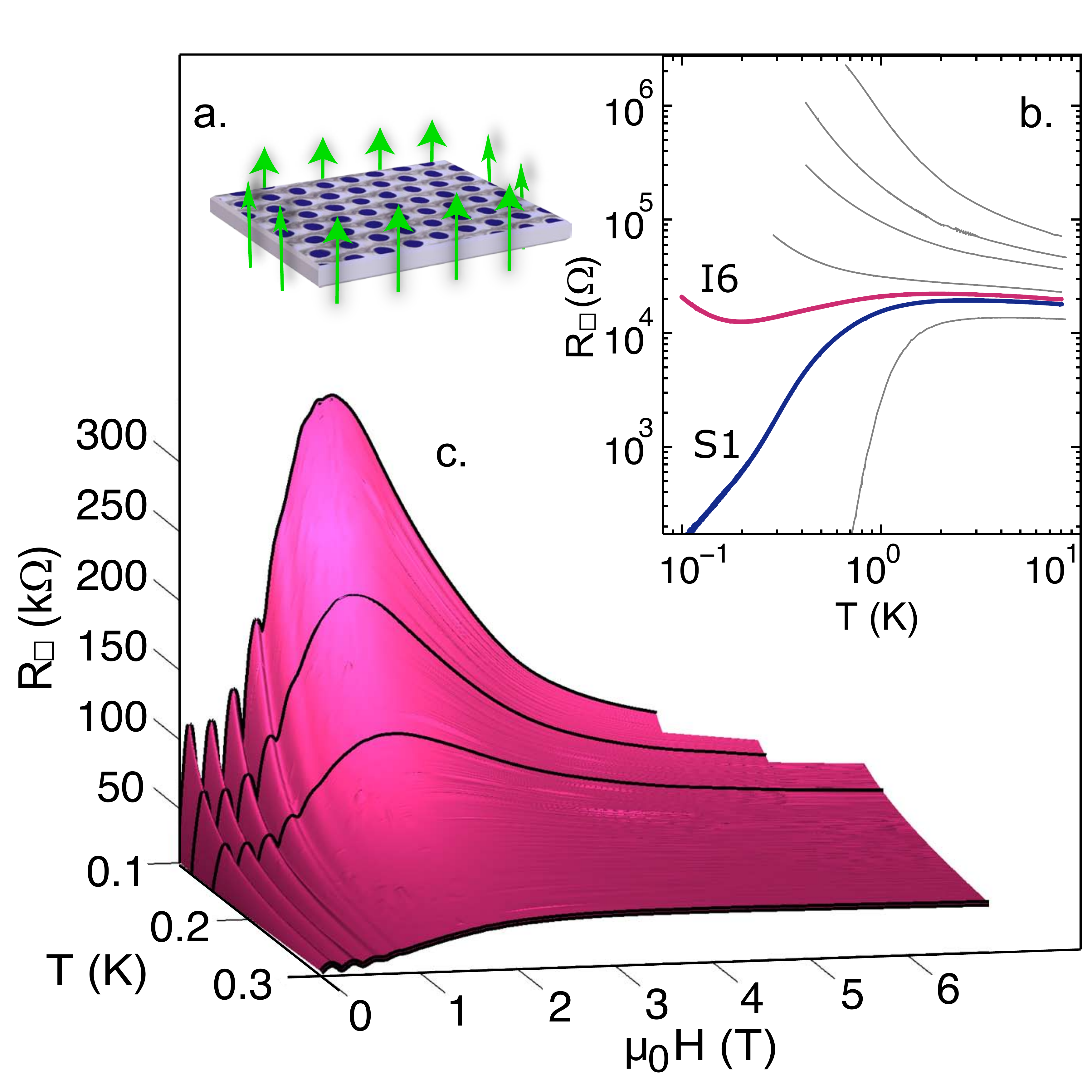}
\caption{a) SEM image of the nano-honeycomb substrate.  The hole center to center spacing and radii are 100 $\pm$ 5 and 27 $\pm$ 3 nm, respectively.  Arrows denote $\vec{H}$. b) Sheet resistance as a function of temperature, $R_\square(T)$, of NHC films produced through a series of Bi evaporations.  The film I6 is the last insulating film and S1 is the first superconducting film in the series.  c) Surface plot of $R_\square(T,H)$ for film I6, which has a normal state sheet resistance of 19.6 k$\Omega$ and 1.1 nm Bi thickness. The solid lines are isotherms.  
\label{cap:3d}}
\end{center}
\end{figure}

Theories of the positive MR peak presume that CPs spontaneously localize into islands or puddles\cite{Fisher:PRL1990,Galitski:PRL2005,Dubi:PRB2006,Fistul:PRL2008}.  On each island, the complex SC order parameter has a well defined amplitude, but electrostatic interactions between islands prevents the development of the long range phase coherence necessary for CP delocalization. A magnetic field induces more phase disorder and localization through the direct coupling of the vector potential to the order parameter phase and by reducing the order parameter amplitude through its CP breaking effects.  At very high $H$, the latter effect leads to a negative MR as the film returns to an unpaired state\cite{DELSING:PRB1994}.  Surprisingly, this basic picture qualitatively accounts for the positive MR peak of smooth InO$_x$ and TiN films.  This agreement and recent STM results\cite{Sacepe:PRL2008}, suggest that a CPI phase can spontaneously develop due to electronic interactions\cite{Ghosal:PRL1998}, $H$\cite{Dubi:Nature2007} and/or disorder induced localization\cite{Feigelman:PRL2007}.  Verifying the connection between the MR peak and the CPI phase requires creating a system exhibiting similar MR features that can be simultaneously probed for the CPI phase.

Here we present the MR of an amorphous film system with a clear CPI phase.  These ultrathin Bi films are patterned with a Nano-HoneyComb (NHC) array of holes\cite{Stewart:Science2007}.  They exhibit a large positive MR that peaks at a field comparable to the estimated average depairing field.  Superimposed on the initial rise of the resistance are oscillations with a period set by the SC flux quantum indicating that localized CPs dominate the transport.   Moreover, the $R_\square(T)$ are activated and the oscillations and the rise in $R_\square(H)$ stem primarily from variations in the activation energy.  At fields well beyond the peak, the transport appears dominated by weakly localized quasiparticles.  Many of these characteristics of the MR of NHC films are shared by unpatterned InO$_x$ and TiN films, which supports earlier proposals that their peaks reflect an underlying CPI phase with transport dominated by CP motion.  Furthermore, the behavior of the activation energy of NHC films is consistent with $H$ controlling the CP localization by tuning the energy characterizing the coupling of the relative phases of the localized CP states.  

The data presented characterize amorphous, NHC Bi films that were deposited {\em{in situ}} and measured in a custom designed dilution refrigerator cryostat\cite{Stewart:PRB2008}.  The films were patterned by depositing them onto aluminum oxide substrates structured with a NHC array of holes (see Fig. \ref{cap:3d})\cite{Yin:APL2001}.  Repeated Bi evaporations yielded a series of films spanning the thickness tuned SIT (see Fig.\ref{cap:3d}b)\cite{Stewart:Science2007}.   Film sheet resistances as a function of $T$ and $H$, $R_\square(T,H)$, were measured using standard four-point AC and DC techniques employing currents sufficiently small to be in the linear regime of the current-voltage characteristics.  A SC solenoid provided $H$ up to 8T transverse to the film plane.   The data presented here come from the two films nearest the SIT, I6 and S1.  Experiments on films on two other substrates yielded very similar results.

Patterning these Bi/Sb films makes it possible to detect the charge of the carriers involved in their transport\cite{Stewart:Science2007} and appears to be essential to the formation of the CPI phase\cite{Stewart:Science2007,Stewart:PhysicaC2009}.   The hole array imposes a spatial period with a unit cell area $S$, which sets the $H$ scale, $H_M=\Phi_0/S$, corresponding to one SC flux quantum, $\Phi_0=h/2e$, per unit cell of the array.  As with Josephson Junction and wire arrays, the thermodynamic and transport properties of NHC films are expected to oscillate with period $H_M$\cite{Tinkham:Intro}.  We will use the frustration, $f=H/H_M$ to specify $H$ with $H_M=0.21$T.

\begin{figure}
\begin{center}
\includegraphics[width=\columnwidth,keepaspectratio]{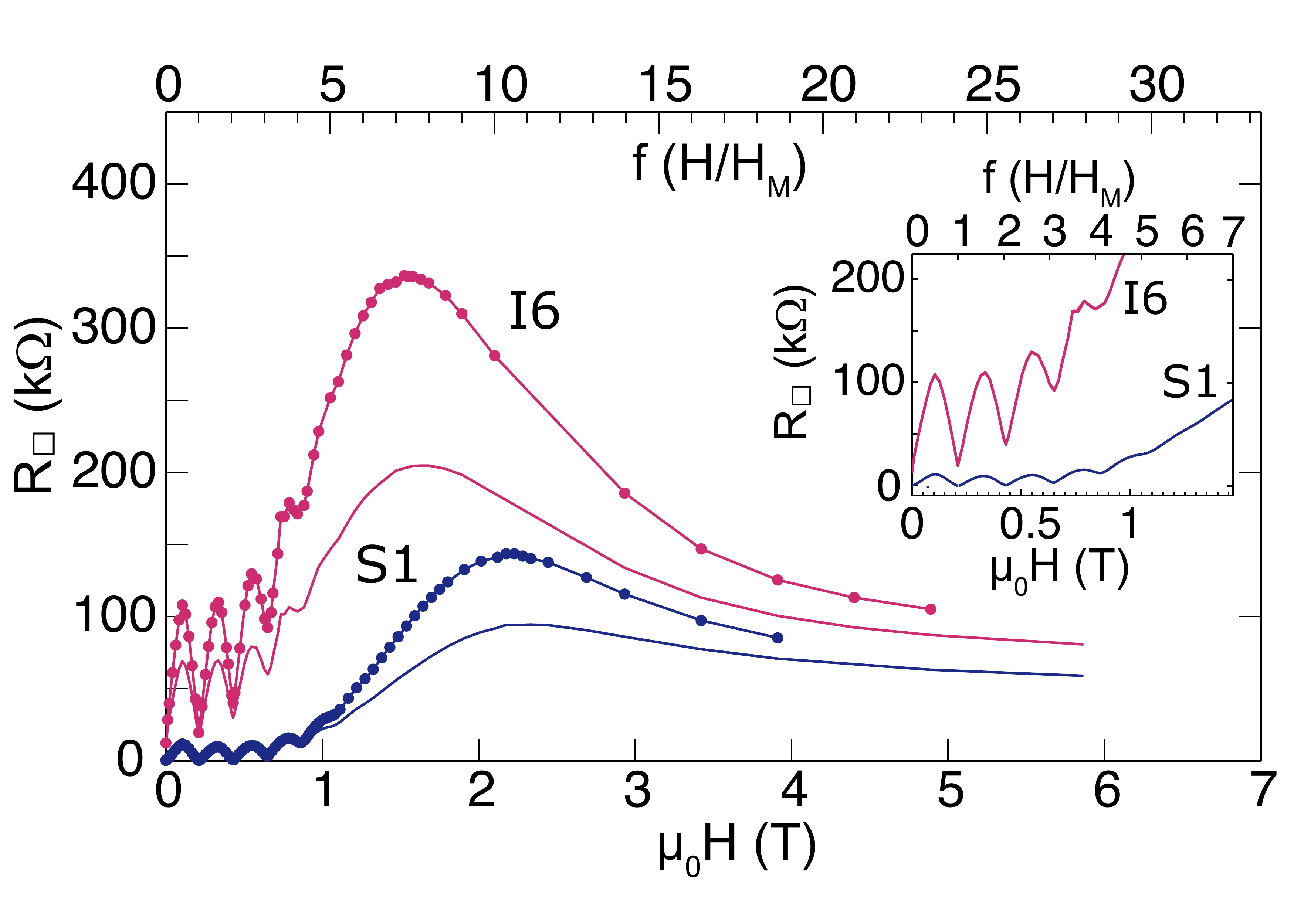}
\caption{Sheet resistance as a function of $H$ at 100 and 120 mK for films I6 and S1.  The lines are spline fits to the data points (shown as symbols on the 100 mK traces).  Inset:  Magnified view of the low $H$ data.  
\label{cap:2film}}
\end{center}
\end{figure}

The MR of NHC films near their thickness-tuned SIT exhibit a rich structure including oscillations at low $H$ that merge into a giant peak at higher $H$.  The $R_\square(T,H)$ surface shown in Fig. \ref{cap:3d} demonstrates these behaviors for the insulating film, I6 (see Fig. \ref{cap:3d}b).  The surface was generated by interpolating data acquired by sweeping $T$ at constant $H$ for a series of closely spaced $H$.  At low $H$ and $T$, the MR oscillates with the period expected for CPs indicating that the film is in the CPI state\cite{Stewart:Science2007}.  The oscillations first appear for $T \simeq$ 0.6K, which gives a lower limit for the CP formation temperature, $T_{c0}$.  At the lowest $T$, five oscillations are easily resolved and there is a hint of a sixth. Oscillations 4-6 are more difficult to resolve because they diminish in amplitude and appear on a rapidly rising background MR.  At 100 mK, $R_\square(H)$ peaks near $f = 7$ with a resistance that is nearly a factor of 30 larger than the zero field value.   This giant MR peak grows and moves to lower field with decreasing $T$.  These $R(T,H)$ features evolve smoothly with decreasing $R_N$ and across the thickness tuned SIT as shown by Fig. \ref{cap:2film} .  The peak moves to higher $H$ and diminishes in size and the oscillation amplitude, measured on a linear scale, shrinks.  We hasten to add that the substrate patterning\cite{Stewart:PhysicaC2009} is essential to the appearance of the MR peak.  Similar Bi/Sb films deposited on glass substrates show only a very small ($<$ 10\%) peak\cite{Jay:thesis}.

\begin{figure}
\begin{center}
\includegraphics[width=\columnwidth,keepaspectratio]{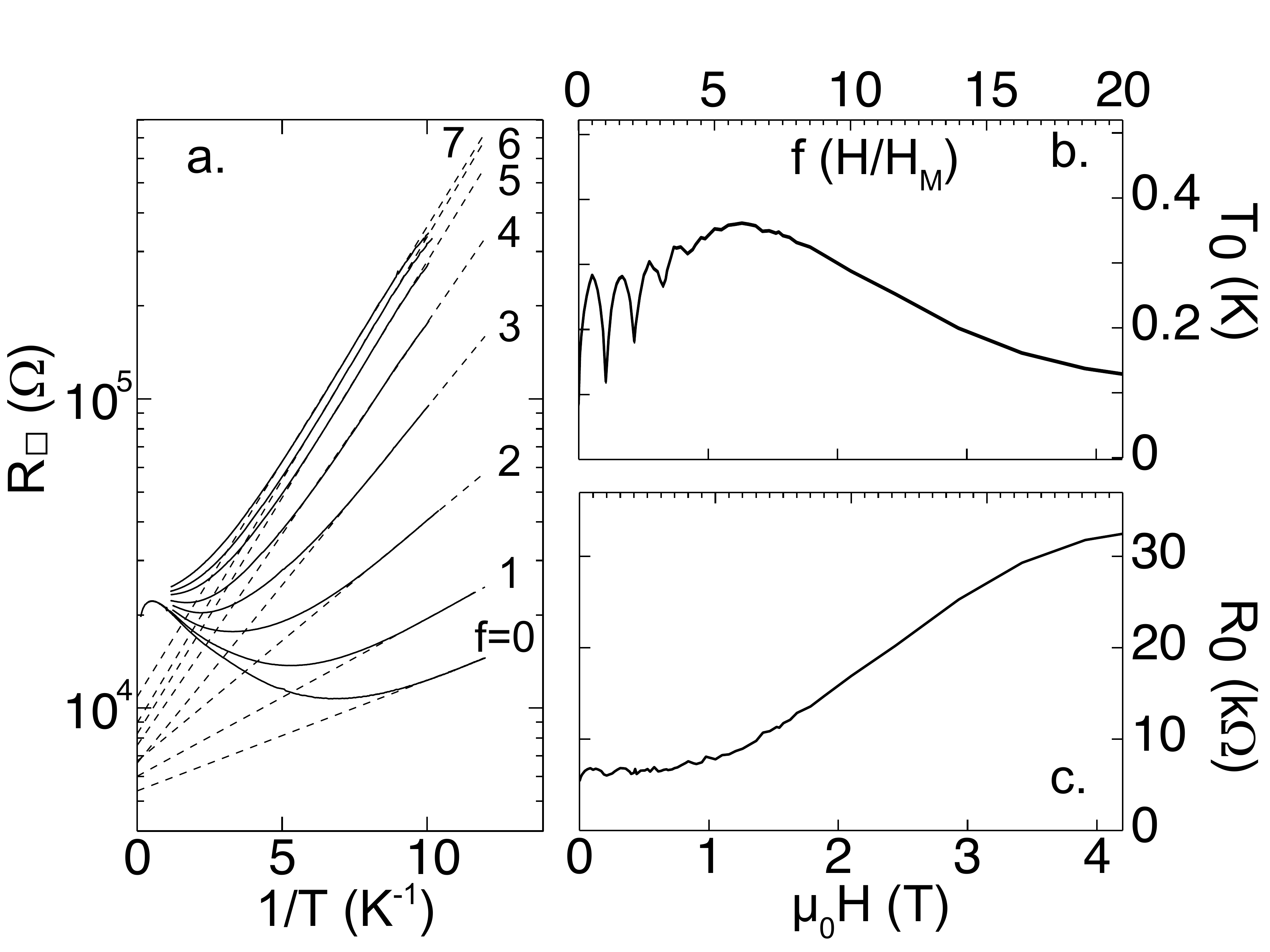}
\caption{ a)  The solid lines are the sheet resistance versus inverse temperature for film I6 in fields corresponding to $f =$ 0,1,2,3,4,5,6,7.  The dashed lines are fits to $R_\square(T)=R_0 exp(T_0/T)$.  b) and c) exhibit the fitting parameters $T_0$ and $R_0$, respectively.   
\label{cap:act}}
\end{center}
\end{figure}

In the low $T$ limit, the $R_\square(T)$ fit well to an Arrhenius form, $R_\square(T)=R_{0}e^{T_{0}/T}$, for $H$ values extending to the peak (see Fig. \ref{cap:act}a).  While only $R_\square(T)$ at integer $f$ are shown, $R_{\square}(T)$ at non-integer $f$ behave similarly.  The activation energy, $T_0$ qualitatively mirrors the MR showing large oscillations with the same period and climbing to a maximum near $f =6$ of $T_0 =0.35 $ K  (see Fig. \ref{cap:act}b).  Moreover, the peak $T_0$ decreases to 0.27 K and moves out to $f\simeq 9$ for the lower $R_N$ film, S1 (not shown).  The prefactor, $R_0$, grows relatively slowly up to the peak field (see Fig. \ref{cap:act}c).  Its factor of 2 increase is substantially smaller than the size of the MR peak at 100 mK.  Taken together, these behaviors indicate that the MR oscillations and the peak in the MR stem primarily from an $H$ dependence of the activation energy. 
 
\begin{figure}
\begin{center}
\includegraphics[width=\columnwidth,keepaspectratio]{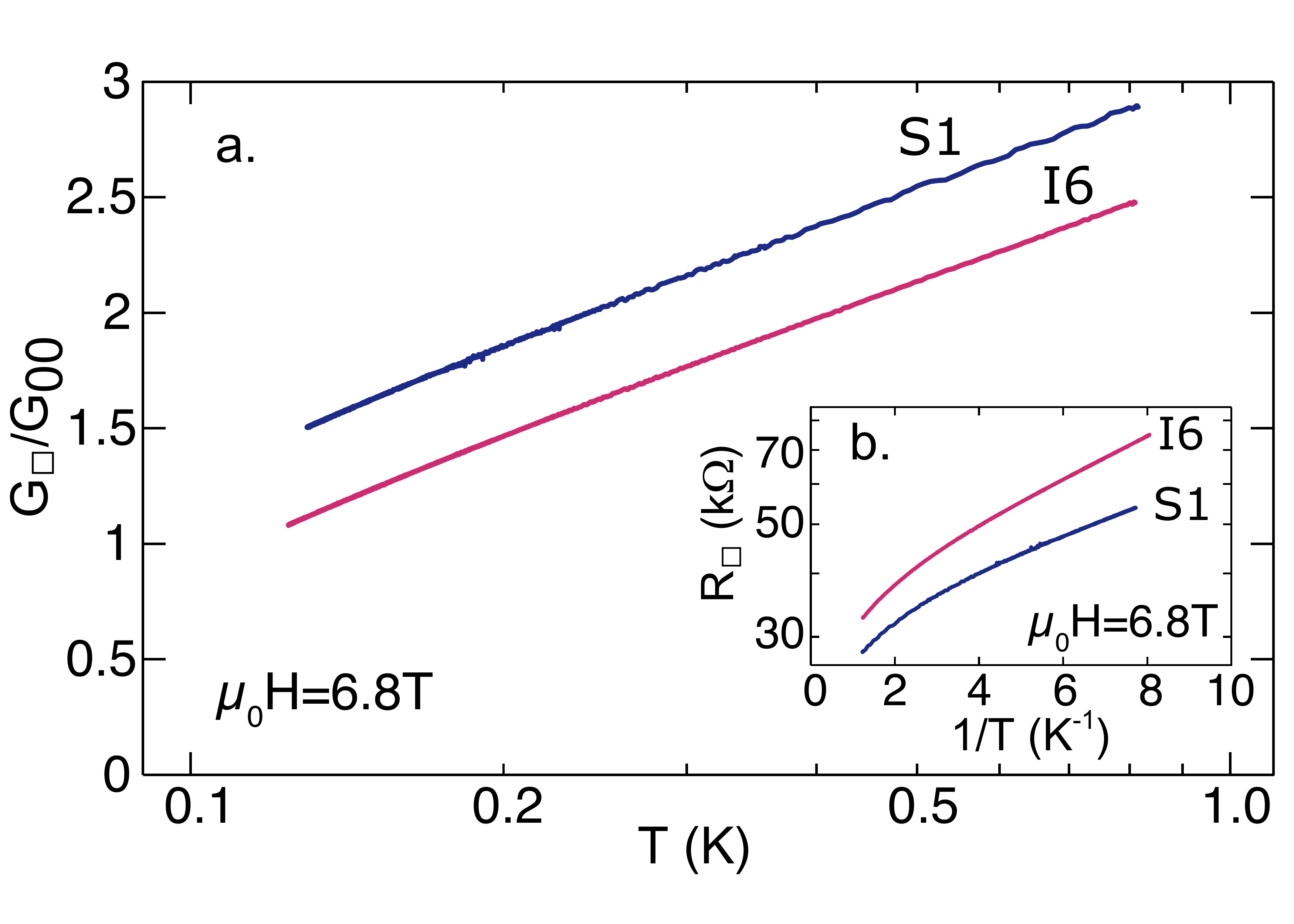}
\caption{Film conductance, $G(T)=1/R_{\square}$ normalized by the quantum of conductance, $G_{00}=e^{2}/2\pi^{2}\hbar$ versus temperature on a logarithmic scale for films I6 and S1 in  $H= 6.8$  T.  Inset:  Arrhenius plots of the same data shown in the main plot. 
\label{cap:wloc}}
\end{center}
\end{figure}

Well beyond the peak ($\mu_0H\gtrsim4$T), where the $R(H)$ appear to asymptotically approach a $T$ dependent constant, the conductance, $G(T)=R^{-1}(T)$ fits better to $G(T)= AG_{00}ln(T/T_0)+G(T_0)$ than an activated form.  Fig. \ref{cap:wloc} compares these forms for  I6 and S1 at 6.8 T. The activated form can only account for the data over a limited range, while the logarithmic dependence extends over nearly a decade. This fit, with $A=0.8\simeq 1$, is consistent with transport by unpaired electrons in the presence of disorder enhanced e-e interactions in two dimensions\cite{BERGMANN:PhysRep1984}.  

The oscillations on the low field side of the MR peak intimate that CP transport processes are at the heart of the giant MR.  For film I6, the oscillations extend to within 20\% of the peak resistance.  Interestingly, S1 displays a similar number of oscillations as I6, but the oscillations do not appear as high on its peak, which appears at a greater field.  This comparison suggests that the number of oscillations is limited by factors such as disorder in the hole array\cite{Benz:PRB1988} rather than the physical effects giving rise to the peak.  Thus, it is possible that CPs exist up to and beyond the peak.

\begin{figure}
\begin{center}
\includegraphics[width=\columnwidth,keepaspectratio]{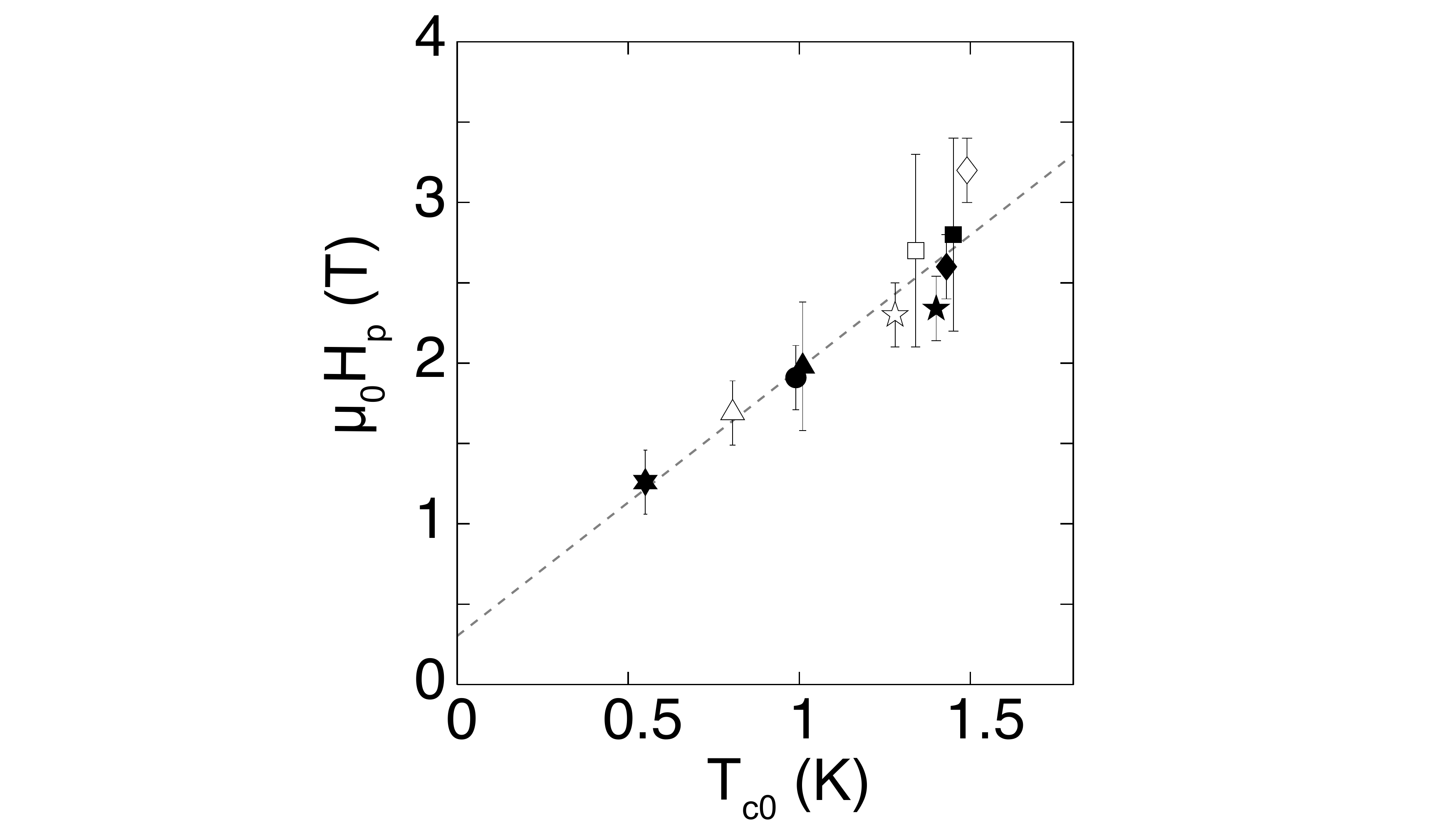}
\caption{ Magnetic field at which the activation energy peaks versus the temperature at which $R(T)$ has dropped to 0.8 of its maximum.  The data points were acquired from films deposited on three different substrates.  Similarly shaped points (open and closed) come from adjacent squares of the same film. Films I6 and S1 are the 6-pointed star and circle, respectively. The dashed line is a weighted linear fit to these data, whose slope is 1.7$\pm$0.2 T/K
and y-intercept=0.3 T.  
\label{cap:Tc0Hp}}
\end{center}
\end{figure}

The values of the peak activation energy, $T_{0}^{max}$, and magnetic field, $H_p$, support the picture of CP dominated transport at low $H$ and the emergence of single electron transport at high $H$.  First,  $T_{0}^{max}$ consistently falls below the CP binding energy, $3.5k_BT_{c0}$, indicating that the low $H$ transport does not involve the breaking of CPs\cite{DELSING:PRB1994}.  Second, $H_p$ appears related to the upper critical $H$, $H_{c2}$, the field required to break CPs\cite{Tinkham:Intro}.  For amorphous films of constant resistivity, dirty limit theory implies, $H_{c2}\propto T_{c0}$\cite{Tinkham:Intro}.  We estimate $T_{c0}$ for the NHC films as the $T$ at which $R(T)$ has dropped to 0.8 of its maximum. We attribute this drop to the onset of very strong pairing fluctuations and fluctuation paraconductivity\cite{Tinkham:Intro}. The choice of 0.8 rather than the standard 0.5 allows analysis of both SC and reentrant films.  Other choices from 0.75 to 0.95 yield similar results for the slope, but cause the intercept to vary.  Data from six films from three different substrates with resistivities estimated to be similar to about 20\% indicate that $H_{p}$ rises approximately linearly with $T_{c0}$ (see Fig.\ref{cap:Tc0Hp}) with a slope of 1.7 $\pm .2$ T/K.  This slope is somewhat larger than what can be estimated from previous experiments on unpatterned films\cite{Chervenak:PRB1996}, which implied that $H_{c2}=1.2T_{c0}$. Thus, $T_{0}^{max}$ appears at or just above the $H$ necessary to depair most of the electrons.
  
These giant MR peaks closely resemble those seen in unpatterned InO$_x$\cite{Sambandamurthy:PRL2004,Steiner:PhysicaC2005,Tan:PRB2008}  and TiN\cite{Baturina:PRL2007,Fistul:PRL2008} films.  Those also grow exponentially with decreasing $T$\cite{Sambandamurthy:PRL2004,Fistul:PRL2008,Steiner:PhysicaC2005,Baturina:PRL2007}, become stronger with disorder, occur near the estimated $H_{c2}$\cite{Steiner:PhysicaC2005}, and evolve smoothly across the disorder tuned SIT\cite{Baturina:PRL2007}.   In addition, their transport is activated and $T_0$ peaks below $T_{c0}$\cite{Sambandamurthy:PRL2004}.  Moreover, at the highest $H$, their $G(T)$ assumes the form for transport by single electrons near the metal to insulator transition\cite{Gantmakher:JETP1998}.  Unlike the NHC films, however, their $G(H)$ appear to saturate at $1/R_Q=e^2/h$, in the zero $T$ limit\cite{Baturina:PRL2007} suggesting a high field quantum metal normal state\cite{Butko:Nature2001} .  This difference could result from stronger spin-orbit scattering in Bi NHC films, which is known to quench the quantum metal state\cite{Xiong:PRB2009}. Overall, the many resemblances intimate that the giant positive MR in the unpatterned films reveals an underlying CPI phase\cite{Dubi:PRB2006,Galitski:PRL2005,Fistul:PRL2008} and the subsequent negative MR reflects the destruction of pairing by $H$.  Previously, the strongest evidence for this association was based on Hall measurements\cite{Paalanen:PRL1992} and scaling analyses supporting the dirty boson picture\cite{Fisher:PRL1990}.   

A notable and perhaps illuminating difference between the NHC and the unpatterned films is the extra structure in the NHC films' $T_0(H)$. Inspection of Fig. \ref{cap:act} suggests that $T_0=T_0^{peak}(H)+T_0^{osc}(H)$, where $T_0^{peak}(H)$ is a peaked function similar to that exhibited by the unpatterned films and $T_0^{osc}(H)$ is a decaying, oscillatory function.  The latter likely originates through extra terms in the Hamiltonian of the form $-E_{J0}cos(\phi_i-\phi_j-2\pi A_{ij}/\Phi_0)$, where the $\phi$'s are phases on neighboring nodes of the array, $A_{ij}$ is the line integral of the vector potential between nodes, and $E_{J0}$ is the zero field phase coupling energy.  Such terms are thought to produce oscillations of the activation energy of insulating micro-fabricated Josephson Junction arrays\cite{DELSING:PRB1994} and can also account for decaying oscillations of $T_{c0}$ of geometrically disordered wire arrays\cite{Benz:PRB1988}.   The absence of oscillations in the unpatterned films does not necessarily imply the absence of a similar Josephson term in their Hamiltonian. Rather, the absence might reflect strong disorder in the geometry of an effective wire array model of the films\cite{Benz:PRB1988,Granato:PRB1986}.The origin of $T_0^{peak}(H)$ is less clear.  The disappearance of oscillations near the peak, the negative MR and the recovery of clear signatures of single electron transport at high fields could indicate the onset of quasiparticle dominated transport at the peak.    Consistently, the close correspondence between $H_p$ and $H_{c2}$ suggests that $T_0^{peak}(H)$'s rise at low fields involves the suppression of the order parameter amplitude. 

In summary, we have shown that the magnetoresistance of amorphous nano-honeycomb Bi films exhibits a giant, positive peak, which is similar to that observed in other unpatterned, ultrathin film systems near the superconductor to insulator transition.  The main result is the verification that this peak arises from the low field positive magnetoresistance of a Cooper Pair Insulator phase with transport dominated by incoherent tunneling of Cooper pairs and a high field negative magnetoresistance associated with the destruction of the pairs.  While these conclusions agree qualitatively with some models, fundamental questions about the processes driving the Cooper pair localization, the emergence of the Cooper Pair Insulator phase at high disorder and the size of the localized states require further experimental and theoretical attention.    

Acknowledgments: We are grateful for helpful conversations with M. V. Feigel'man, D. Feldman, and P. W. Adams.  This work was supported by the NSF through grants No. DMR-0203608 and No. DMR-0605797, by the AFRL, and by the ONR.

\bibliographystyle{apsrev}
\bibliography{MRPeak}

\begin{thebibliography}{31}
\expandafter\ifx\csname natexlab\endcsname\relax\def\natexlab#1{#1}\fi
\expandafter\ifx\csname bibnamefont\endcsname\relax
  \def\bibnamefont#1{#1}\fi
\expandafter\ifx\csname bibfnamefont\endcsname\relax
  \def\bibfnamefont#1{#1}\fi
\expandafter\ifx\csname citenamefont\endcsname\relax
  \def\citenamefont#1{#1}\fi
\expandafter\ifx\csname url\endcsname\relax
  \def\url#1{\texttt{#1}}\fi
\expandafter\ifx\csname urlprefix\endcsname\relax\def\urlprefix{URL }\fi
\providecommand{\bibinfo}[2]{#2}
\providecommand{\eprint}[2][]{\url{#2}}

\bibitem[{\citenamefont{Tinkham}(2004)}]{Tinkham:Intro}
\bibinfo{author}{\bibfnamefont{M.}~\bibnamefont{Tinkham}},
  \emph{\bibinfo{title}{Introduction to Superconductivity}}
  (\bibinfo{publisher}{Courier Dover Publications}, \bibinfo{year}{2004}).

\bibitem[{\citenamefont{Hebard and Paalanen}(1990)}]{Hebard:PRL1990}
\bibinfo{author}{\bibfnamefont{A.~F.} \bibnamefont{Hebard}} \bibnamefont{and}
  \bibinfo{author}{\bibfnamefont{M.~A.} \bibnamefont{Paalanen}},
  \bibinfo{journal}{Phys. Rev. Lett.} \textbf{\bibinfo{volume}{65}},
  \bibinfo{pages}{927} (\bibinfo{year}{1990}).

\bibitem[{\citenamefont{{Gantmakher {\it et al.}}}(1998)}]{Gantmakher:JETP1998}
\bibinfo{author}{\bibfnamefont{V.~F.} \bibnamefont{{Gantmakher {\it et al.}}}},
  \bibinfo{journal}{JETP Lett.} \textbf{\bibinfo{volume}{68}},
  \bibinfo{pages}{363} (\bibinfo{year}{1998}).

\bibitem[{\citenamefont{Sambandamurthy
  et~al.}(2004)\citenamefont{Sambandamurthy, Engel, Johansson, and
  Shahar}}]{Sambandamurthy:PRL2004}
\bibinfo{author}{\bibfnamefont{G.}~\bibnamefont{Sambandamurthy}},
  \bibinfo{author}{\bibfnamefont{L.~W.} \bibnamefont{Engel}},
  \bibinfo{author}{\bibfnamefont{A.}~\bibnamefont{Johansson}},
  \bibnamefont{and} \bibinfo{author}{\bibfnamefont{D.}~\bibnamefont{Shahar}},
  \bibinfo{journal}{Phys. Rev. Lett.} \textbf{\bibinfo{volume}{92}},
  \bibinfo{pages}{107005} (\bibinfo{year}{2004}).

\bibitem[{\citenamefont{{Baturina {\it et al.}}}(2007)}]{Baturina:PRL2007}
\bibinfo{author}{\bibfnamefont{T.~I.} \bibnamefont{{Baturina {\it et al.}}}},
  \bibinfo{journal}{Phys. Rev. Lett.} \textbf{\bibinfo{volume}{99}},
  \bibinfo{pages}{257003} (\bibinfo{year}{2007}).

\bibitem[{\citenamefont{Barber et~al.}(1994)\citenamefont{Barber, Merchant,
  LaPorta, and Dynes}}]{Barber:PRB1994}
\bibinfo{author}{\bibfnamefont{R.~P.} \bibnamefont{Barber}},
  \bibinfo{author}{\bibfnamefont{L.~M.} \bibnamefont{Merchant}},
  \bibinfo{author}{\bibfnamefont{A.}~\bibnamefont{LaPorta}}, \bibnamefont{and}
  \bibinfo{author}{\bibfnamefont{R.~C.} \bibnamefont{Dynes}},
  \bibinfo{journal}{Phys. Rev. B} \textbf{\bibinfo{volume}{49}},
  \bibinfo{pages}{3409} (\bibinfo{year}{1994}).

\bibitem[{\citenamefont{{Stewart Jr.} et~al.}(2007)\citenamefont{{Stewart Jr.},
  Yin, Xu, and {Valles Jr.}}}]{Stewart:Science2007}
\bibinfo{author}{\bibfnamefont{M.~D.} \bibnamefont{{Stewart Jr.}}},
  \bibinfo{author}{\bibfnamefont{A.}~\bibnamefont{Yin}},
  \bibinfo{author}{\bibfnamefont{J.~M.} \bibnamefont{Xu}}, \bibnamefont{and}
  \bibinfo{author}{\bibfnamefont{J.~M.} \bibnamefont{{Valles Jr.}}},
  \bibinfo{journal}{Science} \textbf{\bibinfo{volume}{318}},
  \bibinfo{pages}{1273} (\bibinfo{year}{2007}).

\bibitem[{\citenamefont{{Stewart Jr.} et~al.}(2008)\citenamefont{{Stewart Jr.},
  Yin, Xu, and {Valles Jr.}}}]{Stewart:PRB2008}
\bibinfo{author}{\bibfnamefont{M.~D.} \bibnamefont{{Stewart Jr.}}},
  \bibinfo{author}{\bibfnamefont{A.}~\bibnamefont{Yin}},
  \bibinfo{author}{\bibfnamefont{J.~M.} \bibnamefont{Xu}}, \bibnamefont{and}
  \bibinfo{author}{\bibfnamefont{J.~M.} \bibnamefont{{Valles Jr.}}},
  \bibinfo{journal}{Phys. Rev. B} \textbf{\bibinfo{volume}{77}},
  \bibinfo{pages}{140501(R)} (\bibinfo{year}{2008}).

\bibitem[{\citenamefont{Ekinci and {Valles Jr.}}(1999)}]{Ekinci:PRL1999}
\bibinfo{author}{\bibfnamefont{K.~L.} \bibnamefont{Ekinci}} \bibnamefont{and}
  \bibinfo{author}{\bibfnamefont{J.~M.} \bibnamefont{{Valles Jr.}}},
  \bibinfo{journal}{Phys. Rev. Lett.} \textbf{\bibinfo{volume}{82}},
  \bibinfo{pages}{1518} (\bibinfo{year}{1999}).

\bibitem[{\citenamefont{{Barber Jr.} et~al.}(2006)\citenamefont{{Barber Jr.},
  Hsu, {Valles Jr.}, Dynes, and Glover}}]{Barber:PRB2006}
\bibinfo{author}{\bibfnamefont{R.~P.} \bibnamefont{{Barber Jr.}}},
  \bibinfo{author}{\bibfnamefont{S.~Y.} \bibnamefont{Hsu}},
  \bibinfo{author}{\bibfnamefont{J.~M.} \bibnamefont{{Valles Jr.}}},
  \bibinfo{author}{\bibfnamefont{R.~C.} \bibnamefont{Dynes}}, \bibnamefont{and}
  \bibinfo{author}{\bibfnamefont{R.~E.} \bibnamefont{Glover}},
  \bibinfo{journal}{Phys. Rev. B} \textbf{\bibinfo{volume}{73}},
  \bibinfo{pages}{134516} (\bibinfo{year}{2006}).

\bibitem[{\citenamefont{Steiner and Kapitulnik}(2005)}]{Steiner:PhysicaC2005}
\bibinfo{author}{\bibfnamefont{M.~A.} \bibnamefont{Steiner}} \bibnamefont{and}
  \bibinfo{author}{\bibfnamefont{A.}~\bibnamefont{Kapitulnik}},
  \bibinfo{journal}{Physica C} \textbf{\bibinfo{volume}{422}},
  \bibinfo{pages}{16} (\bibinfo{year}{2005}).

\bibitem[{\citenamefont{{Sarwa B. Tan} et~al.}(2008)\citenamefont{{Sarwa B.
  Tan}, Parendo, and Goldman}}]{Tan:PRB2008}
\bibinfo{author}{\bibfnamefont{K.~H.} \bibnamefont{{Sarwa B. Tan}}},
  \bibinfo{author}{\bibfnamefont{K.~A.} \bibnamefont{Parendo}},
  \bibnamefont{and} \bibinfo{author}{\bibfnamefont{A.~M.}
  \bibnamefont{Goldman}}, \bibinfo{journal}{Phys. Rev. B}
  \textbf{\bibinfo{volume}{78}}, \bibinfo{pages}{014506}
  (\bibinfo{year}{2008}).

\bibitem[{\citenamefont{Fisher}(1990)}]{Fisher:PRL1990}
\bibinfo{author}{\bibfnamefont{M.~P.~A.} \bibnamefont{Fisher}},
  \bibinfo{journal}{Phys. Rev. Lett.} \textbf{\bibinfo{volume}{65}},
  \bibinfo{pages}{923} (\bibinfo{year}{1990}).

\bibitem[{\citenamefont{Galitski et~al.}(2005)\citenamefont{Galitski, Refael,
  Fisher, and Senthil}}]{Galitski:PRL2005}
\bibinfo{author}{\bibfnamefont{V.~M.} \bibnamefont{Galitski}},
  \bibinfo{author}{\bibfnamefont{G.}~\bibnamefont{Refael}},
  \bibinfo{author}{\bibfnamefont{M.~P.~A.} \bibnamefont{Fisher}},
  \bibnamefont{and} \bibinfo{author}{\bibfnamefont{T.}~\bibnamefont{Senthil}},
  \bibinfo{journal}{Phys. Rev. Lett.} \textbf{\bibinfo{volume}{95}},
  \bibinfo{pages}{077002} (\bibinfo{year}{2005}).

\bibitem[{\citenamefont{Dubi et~al.}(2006)\citenamefont{Dubi, Meir, and
  Avishai}}]{Dubi:PRB2006}
\bibinfo{author}{\bibfnamefont{Y.}~\bibnamefont{Dubi}},
  \bibinfo{author}{\bibfnamefont{Y.}~\bibnamefont{Meir}}, \bibnamefont{and}
  \bibinfo{author}{\bibfnamefont{Y.}~\bibnamefont{Avishai}},
  \bibinfo{journal}{Phys. Rev. B} \textbf{\bibinfo{volume}{73}},
  \bibinfo{pages}{054509} (\bibinfo{year}{2006}).

\bibitem[{\citenamefont{Fistul et~al.}(2008)\citenamefont{Fistul, Vinokur, and
  Baturina}}]{Fistul:PRL2008}
\bibinfo{author}{\bibfnamefont{M.~V.} \bibnamefont{Fistul}},
  \bibinfo{author}{\bibfnamefont{V.~M.} \bibnamefont{Vinokur}},
  \bibnamefont{and} \bibinfo{author}{\bibfnamefont{T.~I.}
  \bibnamefont{Baturina}}, \bibinfo{journal}{Phys. Rev. Lett.}
  \textbf{\bibinfo{volume}{100}}, \bibinfo{pages}{086805}
  (\bibinfo{year}{2008}).

\bibitem[{\citenamefont{{Delsing {\it et al.}}}(1994)}]{DELSING:PRB1994}
\bibinfo{author}{\bibfnamefont{P.}~\bibnamefont{{Delsing {\it et al.}}}},
  \bibinfo{journal}{Phys. Rev. B} \textbf{\bibinfo{volume}{50}},
  \bibinfo{pages}{3959} (\bibinfo{year}{1994}).

\bibitem[{\citenamefont{{Sacepe {\it et al.}}}(2008)}]{Sacepe:PRL2008}
\bibinfo{author}{\bibfnamefont{B.}~\bibnamefont{{Sacepe {\it et al.}}}},
  \bibinfo{journal}{Phys. Rev. Lett.} \textbf{\bibinfo{volume}{101}},
  \bibinfo{pages}{157006} (\bibinfo{year}{2008}).

\bibitem[{\citenamefont{Ghosal et~al.}(1998)\citenamefont{Ghosal, Randeria, and
  Trivedi}}]{Ghosal:PRL1998}
\bibinfo{author}{\bibfnamefont{A.}~\bibnamefont{Ghosal}},
  \bibinfo{author}{\bibfnamefont{M.}~\bibnamefont{Randeria}}, \bibnamefont{and}
  \bibinfo{author}{\bibfnamefont{N.}~\bibnamefont{Trivedi}},
  \bibinfo{journal}{Phys. Rev. Lett.} \textbf{\bibinfo{volume}{81}},
  \bibinfo{pages}{3940} (\bibinfo{year}{1998}).

\bibitem[{\citenamefont{Dubi et~al.}(2007)\citenamefont{Dubi, Meir, and
  Avishai}}]{Dubi:Nature2007}
\bibinfo{author}{\bibfnamefont{Y.}~\bibnamefont{Dubi}},
  \bibinfo{author}{\bibfnamefont{Y.}~\bibnamefont{Meir}}, \bibnamefont{and}
  \bibinfo{author}{\bibfnamefont{Y.}~\bibnamefont{Avishai}},
  \bibinfo{journal}{Nature} \textbf{\bibinfo{volume}{449}},
  \bibinfo{pages}{876} (\bibinfo{year}{2007}).

\bibitem[{\citenamefont{Feigel'man et~al.}(2007)\citenamefont{Feigel'man,
  Ioffe, Kravtsov, and Yuzbashyan}}]{Feigelman:PRL2007}
\bibinfo{author}{\bibfnamefont{M.~V.} \bibnamefont{Feigel'man}},
  \bibinfo{author}{\bibfnamefont{L.~B.} \bibnamefont{Ioffe}},
  \bibinfo{author}{\bibfnamefont{V.~E.} \bibnamefont{Kravtsov}},
  \bibnamefont{and} \bibinfo{author}{\bibfnamefont{E.~A.}
  \bibnamefont{Yuzbashyan}}, \bibinfo{journal}{Phys. Rev. Lett.}
  \textbf{\bibinfo{volume}{98}}, \bibinfo{pages}{027001}
  (\bibinfo{year}{2007}).

\bibitem[{\citenamefont{{Yin {\it et al.}}}(2001)}]{Yin:APL2001}
\bibinfo{author}{\bibfnamefont{A.}~\bibnamefont{{Yin {\it et al.}}}},
  \bibinfo{journal}{Appl. Phys. Lett.} \textbf{\bibinfo{volume}{79}},
  \bibinfo{pages}{1039} (\bibinfo{year}{2001}).

\bibitem[{\citenamefont{{Stewart Jr.} et~al.}(2009)\citenamefont{{Stewart Jr.},
  Nguyen, Hollen, Yin, Xu, and {Valles Jr.}}}]{Stewart:PhysicaC2009}
\bibinfo{author}{\bibfnamefont{M.~D.} \bibnamefont{{Stewart Jr.}}},
  \bibinfo{author}{\bibfnamefont{H.~Q.} \bibnamefont{Nguyen}},
  \bibinfo{author}{\bibfnamefont{S.~M.} \bibnamefont{Hollen}},
  \bibinfo{author}{\bibfnamefont{A.}~\bibnamefont{Yin}},
  \bibinfo{author}{\bibfnamefont{J.~M.} \bibnamefont{Xu}}, \bibnamefont{and}
  \bibinfo{author}{\bibfnamefont{J.~M.} \bibnamefont{{Valles Jr.}}},
  \bibinfo{journal}{Physica C} \textbf{\bibinfo{volume}{469}},
  \bibinfo{pages}{774} (\bibinfo{year}{2009}).

\bibitem[{\citenamefont{Chervenak}(1998)}]{Jay:thesis}
\bibinfo{author}{\bibfnamefont{J.~A.} \bibnamefont{Chervenak}}, Ph.D. thesis,
  \bibinfo{school}{Brown University, Providence, RI} (\bibinfo{year}{1998}).

\bibitem[{\citenamefont{Bergmann}(1984)}]{BERGMANN:PhysRep1984}
\bibinfo{author}{\bibfnamefont{G.}~\bibnamefont{Bergmann}},
  \bibinfo{journal}{Phys. Rep.} \textbf{\bibinfo{volume}{107}},
  \bibinfo{pages}{1} (\bibinfo{year}{1984}).

\bibitem[{\citenamefont{Benz et~al.}(1988)\citenamefont{Benz, Forrester,
  Tinkham, and Lobb}}]{Benz:PRB1988}
\bibinfo{author}{\bibfnamefont{S.~P.} \bibnamefont{Benz}},
  \bibinfo{author}{\bibfnamefont{M.~G.} \bibnamefont{Forrester}},
  \bibinfo{author}{\bibfnamefont{M.}~\bibnamefont{Tinkham}}, \bibnamefont{and}
  \bibinfo{author}{\bibfnamefont{C.~J.} \bibnamefont{Lobb}},
  \bibinfo{journal}{Phys. Rev. B} \textbf{\bibinfo{volume}{38}},
  \bibinfo{pages}{2869} (\bibinfo{year}{1988}).

\bibitem[{\citenamefont{Chervenak and {Valles Jr.}}(1996)}]{Chervenak:PRB1996}
\bibinfo{author}{\bibfnamefont{J.~A.} \bibnamefont{Chervenak}}
  \bibnamefont{and} \bibinfo{author}{\bibfnamefont{J.~M.} \bibnamefont{{Valles
  Jr.}}}, \bibinfo{journal}{Phys. Rev. B} \textbf{\bibinfo{volume}{54}},
  \bibinfo{pages}{R15649} (\bibinfo{year}{1996}).

\bibitem[{\citenamefont{Butko and Adams}(2001)}]{Butko:Nature2001}
\bibinfo{author}{\bibfnamefont{V.~Y.} \bibnamefont{Butko}} \bibnamefont{and}
  \bibinfo{author}{\bibfnamefont{P.~W.} \bibnamefont{Adams}},
  \bibinfo{journal}{Nature} \textbf{\bibinfo{volume}{409}},
  \bibinfo{pages}{161} (\bibinfo{year}{2001}).

\bibitem[{\citenamefont{Xiong et~al.}(2009)\citenamefont{Xiong, Karki, Young,
  and Adams}}]{Xiong:PRB2009}
\bibinfo{author}{\bibfnamefont{Y.~M.} \bibnamefont{Xiong}},
  \bibinfo{author}{\bibfnamefont{A.~B.} \bibnamefont{Karki}},
  \bibinfo{author}{\bibfnamefont{D.~P.} \bibnamefont{Young}}, \bibnamefont{and}
  \bibinfo{author}{\bibfnamefont{P.~W.} \bibnamefont{Adams}},
  \bibinfo{journal}{Phys. Rev. B} \textbf{\bibinfo{volume}{79}},
  \bibinfo{pages}{020510(R)} (\bibinfo{year}{2009}).

\bibitem[{\citenamefont{Paalanen et~al.}(1992)\citenamefont{Paalanen, Hebard,
  and Ruel}}]{Paalanen:PRL1992}
\bibinfo{author}{\bibfnamefont{M.~A.} \bibnamefont{Paalanen}},
  \bibinfo{author}{\bibfnamefont{A.~F.} \bibnamefont{Hebard}},
  \bibnamefont{and} \bibinfo{author}{\bibfnamefont{R.~R.} \bibnamefont{Ruel}},
  \bibinfo{journal}{Phys Rev Lett} \textbf{\bibinfo{volume}{69}},
  \bibinfo{pages}{1604} (\bibinfo{year}{1992}).

\bibitem[{\citenamefont{Granato and Kosterlitz}(1986)}]{Granato:PRB1986}
\bibinfo{author}{\bibfnamefont{E.}~\bibnamefont{Granato}} \bibnamefont{and}
  \bibinfo{author}{\bibfnamefont{J.~M.} \bibnamefont{Kosterlitz}},
  \bibinfo{journal}{Phys. Rev. B} \textbf{\bibinfo{volume}{33}},
  \bibinfo{pages}{6533} (\bibinfo{year}{1986}).

\end{thebibliography}

\end{document}